# Exploring the near-surface at the lunar South Pole with geophysical tools

C. Schmelzbach₁, S. Stähler₁, N. C. Schmerr₂, M. Knapmeyer₃, D. Sollberger₁, P. Edme₁, A. Khan₁, N. Brinkman₁, L. Ferraioli₁, J. O. A. Robertsson₁, D. Giardini₁

₁ETH Zurich, Switzerland; ₂University of Maryland, College Park, USA; ₃DLR, Berlin, Germany

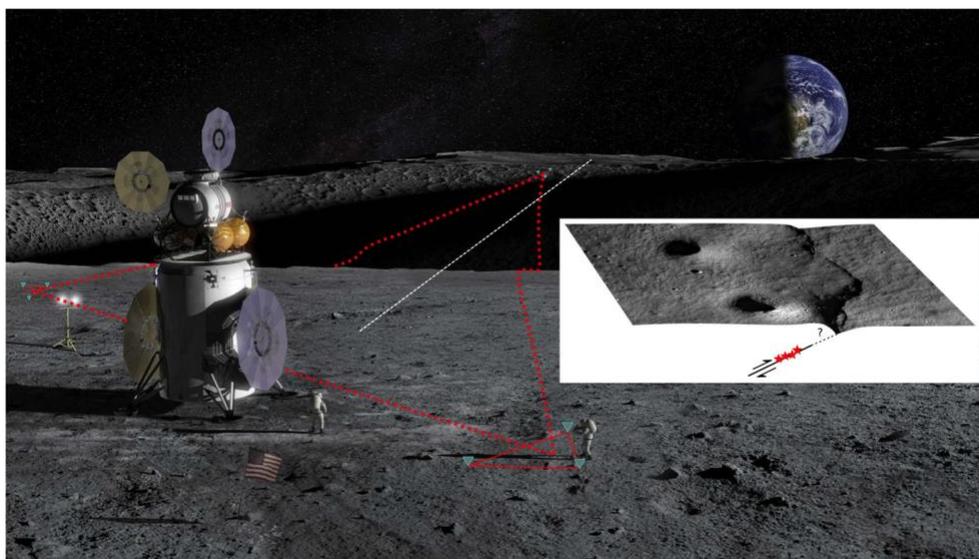

*Figure 1. Conceptual view of the Artemis III lander with three seismic mini-arrays (triangles) connected by fibre-optic cables (red dashed lines). Inset shows a lobate scarp with potential moonquakes indicated by red stars (LROC NAC M159099396R, NASA/GSFC/Arizona State University).*

## MOTIVATION

The Moon is a natural wonder and a potential host of resources that will one day enable human exploration of the Solar System [1]. Imaging, exploring and understanding the lunar near-surface structure (topmost few hundreds of meters) and processes will be key for *in situ* resource utilization (ISRU), identification of hazards for crews and infrastructure, and answering science questions on the formation and interior of the Moon. Compared to the equatorial plains shaped by perennial solar wind and repeated impact bombardment, the lunar poles have likely a higher abundance of volatiles. The poles are therefore expected to have a a significantly different near-surface zone than for example the sites visited by the Apollo missions.

Near-surface investigations of a crewed mission like *Artemis III* at the lunar South Pole would address key science questions: What natural resources of importance for ISRU are available (e.g., pure water ice, regolith building materials, void spaces/caves as shelter)? What are the physical properties of the polar regolith as a host of resources and as a resource by itself? What is the internal structure of impact craters, basins, and other features like lobate scarps relevant to resolve the past and recent geological and tectonic history? Answering these questions will be vital to establishing permanent outposts on the Moon as well as to investigate the history of the Moon and the inner Solar System.

The Moon was studied using a suite of geophysical experiments during the Apollo missions including active and passive seismic measurements [2], heat flow [3], magnetic [4], electromagnetic [5] and gravimetric [6] soundings, and the installation of retroreflectors for ranging experiments [7]. These investigations informed on the near-surface regolith structure, indicated the presence of a lunar megaregolith, and provided a picture of the deep lunar crust, mantle and core [8].

The *Artemis III* mission provides an opportunity to build and expand upon the key success of Apollo in studying the lunar near-surface environment, and to take advantage of the significant developments in geophysical instrumentation, data acquisition and experimental design, as well as data analysis that has taken place since that era. *The goal of this white paper is to highlight the value of ground-based geophysical experiments by a crew and to propose a series of experiments to address key science questions.*

## GEOPHYSICAL IMAGING OF THE COMPLEX LUNAR NEAR SURFACE IN A POLAR REGION

The lunar near-surface environment is expected to have a complex composition. For example, much of the scattering effects observed on the Apollo seismic data are thought to be due to the shallowest layers of the regolith [9, 10] from the poorly mixed presence of ejecta materials and brecciated lunar bedrock. As a consequence of the near-surface heterogeneity, the interpretation of geophysical measurements is often non-unique. Multi-disciplinary approaches including different geophysical methods as well as geological and engineering techniques are thus necessary. Data from different measurements can be jointly interpreted to constrain near-surface geophysical model building [11].

The initial *Artemis III* mission to the lunar South Pole will only allow probing one or a few sites within several 10's to 100's of meters of the landing site. The constraints imposed by space missions in terms of instrument weight, data transfer volume, number and location of measurement points require approaches that extract the maximum possible information from a few measurements only. Novel ideas of sampling [12] and extracting additional observables from conventional equipment such as spatial gradient information from mini-arrays of instruments [13] are paramount. For example, Sollberger et al. [14] demonstrate the value of latest data processing techniques when applied to the Apollo active seismic dataset, but also show that new experiments can be designed with such techniques in mind to enable to move beyond the results of the Apollo age.

Jointly analysing data from several geophysical exploration techniques acquired in combined human-operated and robotic activities and supported by other ground-based and orbital observations will be key for successful near-surface investigations. Wavefield-based techniques such as seismic and ground penetrating radar (GPR) exploration have proven to be useful tools for shallow subsurface imaging on the Moon during the Apollo [15] and more recent Chang'e 3 and 4 missions [16], respectively, to provide high-resolution images and physical property models. As seismic and GPR imaging respond to elastic and electric properties, respectively, they can provide complementary subsurface models. Diffusive-field methods such as electro-magnetic (EM) and electric sounding allow mapping the subsurface electrical resistivity structure and may be able to probe the lunar subsurface at great depth due to the electrically very resistive lunar environment [17]. The interpretation of potential field techniques such as gravimetry and magnetometry provides further insights into subsurface structures at different scales but often requires other data as constraints. While most of the techniques listed above are active methods used for profiling, it will be critical to complement these activities with passive monitoring installations (e.g., seismometers, heat flow, EM sounding), ideally leveraging the mutual benefit when linked with a future Lunar Geophysical Network [18].

### TARGETS FOR GEOPHYSICAL INVESTIGATIONS AT THE LUNAR SOUTH POLE

In the following, we outline a series of geophysical experiments that could be carried out by the *Artemis III* astronauts to explore the area around the landing site.

**I) A lunar fault monitoring observatory across a lobate scarp**

The recent discovery of young thrust faults and a potential link between these faults and shallow moonquakes (SMQ) observed on the Apollo seismometers could indicate recent or even on-going tectonic activity on the Moon [19-21]. Such tectonic activity is not only of significant scientific interest providing a window into the recent lunar stress state but would also represent a potential hazard for human activities (one of the largest SMQ with moment magnitude ~4 occurred ~6° from the South Pole [22]).

We propose to combine *in situ* geophysical investigations with seismic monitoring to investigate one of the ubiquitous lobate scarps [21]. Seismic and/or GPR profiling across a lobate scarp has the potential to image structures within the topmost tens to few hundreds of meters to resolve the history of the fault and infer on the level of past and present tectonic activity (Fig. 1).

In addition, we propose to emplace in the vicinity to the lobate scarp a geophysical node with a broad-band seismometer connected to the *Artemis III* lander with a fibre-optic cable that could be used, at the same time, as communication means and a light-weight seismic array of unprecedently high spatial resolution [23]. The node will also include a mini-array of three or four geophones (Fig. 1). Such configurations will allow computing spatial wavefield gradients of the seismic wavefield that greatly enhance the analysis of spatial and temporal seismicity patterns. Establishing such an observatory would be a major step forward in understanding lunar tectonic activity as an example of tectonics of a planetary body with single, thick lithosphere (NRC concepts 2 and 3 [1]).

**II) Physical properties of the regolith at the landing site**

The physical properties of the lunar polar regolith are unknown, but are critical to assess the stability and safety of infrastructures, the utilization of regolith as building material, the extraction of volatiles, trafficability as well as to study regolith processes, weathering on anhydrous airless bodies, and volatile sources and cycles (NRC concepts 4 and 7 [1]).

With the goal to resolve the physical and geotechnical regolith properties at and around the landing site, we propose to install a mini-array of geophones (see Fig. 1). Combined active-source (controlled by astronauts; similarly to the Apollo seismic experiments [15]), artificial impacts (e.g., final stage of *Artemis III* ascent vehicle) and meteorite impacts will enable constructing a near-surface elastic property model (e.g., Poisson's ratio [14] and (dynamic) shear modulus). Complementary GPR and EM measurements would provide high resolution images of the subsurface structure and physical parameter models to, for example, calibrate orbital radar measurements.

**III) Structure and *in-situ* properties of permanently shadowed region – search for water ice and other cold-trapped volatiles**

Lunar water ice and other cold-trapped volatiles are of significant interest not only as scientific repositories for understanding the evolution of the Solar System but also for ISRU.

While the existence of pure ice deposits in permanently shadowed regions (PSR) has been proposed based on Chandraayan-1 and LRO radar data [24, 25], *in-situ* measurements are needed to determine the exact nature and extend of ice and other cold-trapped volatiles at or in the lunar near-surface (NRC concepts 4 and 7 [1]).

We propose to investigate PSRs with multidisciplinary geophysical experiments to address science questions like: What is the water ice content in the shallow subsurface? Is the water ice dispersed through the regolith or concentrated in bodies? Seismic, GPR and EM methods will likely exhibit different sensitivity to subsurface structure and ice content. For example, extended ice bodies likely exhibit significantly higher seismic velocities than shallow regolith layers (up to 3800 m/s for water ice vs. few 100's m/s for regolith observed at the Apollo landing sites), but electric parameter contrasts between regolith and ice bodies may be small. In contrast, GPR images can provide very detailed images of regolith structures as well as aid in the interpretation of surface samples and the extrapolation of ground-based chemical and physical measurements.

**IV) Imaging the interior of the South-Pole Aitken basin**

Landing at the South Pole offers the unique opportunity to investigate the age and structure of the South-Pole Aitken (SPA) basin, which is the largest (~2600 km diameter) and oldest known impact structure in the Solar System [26]. Seismic experiments (for example carried out at a few selected sites), GPR profiling and EM, gravimetric and magnetic measurements along transects will over a unique opportunity to resolve the shallow subsurface structure of the SPA basin, where lunar lower crust and mantle may have been excavated. Geophysical subsurface images will complement orbital observations, geological field work and sample collection. The geophysical models will be critical to test hypotheses on, for example, the SPA formation (NRC concept 1 [1]).

**Implementation**

We expect that any landing site chosen to fill the NRC report concepts [1] will be suitable to carry out the experiments detailed in this white paper [27]. The geophysical soundings and installation of the monitoring equipment will involve a limited number of instruments. Optimizing the experimental layouts and minimizing the necessary time for human control by exploiting latest survey design ideas [12] will allow minimizing the crew time.

### OUTLOOK: EXPLORATION BEYOND THE MOON

The Moon will serve as a comprehensive testbed for extra-terrestrial geophysics. Hence, lessons learned from human geophysical exploration of the Moon as proposed here will be key for the exploration of the moons of Mars and space beyond the Moon, and prepare us for the human exploration of space beyond the Moon.